\documentclass[aps,pra,twocolumn,showpacs,preprintnumbers,amsmath,amssymb]{revtex4}
\usepackage{graphicx}
\usepackage{dcolumn}
\usepackage{bm}

\begin{document}

\title{Interference-induced splitting of resonances in spontaneous emission}
\author{R. Arun}
\email{arunprl@yahoo.com}
\affiliation{Physical Research Laboratory, Navrangpura, Ahmedabad 380 009, India.}

\date{\today}

\begin{abstract}
We study the resonance fluorescence from a coherently driven four-level
atom in the Y-type configuration. The effects of quantum interference 
induced by spontaneous emission on the fluorescence properties of the atom 
are investigated. It is found that the quantum interference resulting from 
cascade emission decays of the atom leads to a splitting of resonances in the 
excited level populations calculated as a function of light detuning. 
For some parameters, interference assisted enhancement of inner sidebands and narrowing of 
central peaks may also occur in the fluorescence spectrum. We present a physical understanding
of our numerical results using the dressed state description of the atom-light interaction.
\end{abstract}

\pacs{42.50.Ct,42.25.Hz,32.50.+d}

\maketitle

\newpage
\section{Introduction}
The study of quantum interference effects in the spontaneous emission of excited atoms
has attracted substantial attention in the literature \cite{{gsa},{zhu1},{zhu2},{pas1},
{pas2},{cardi},{zhou1},{zhou2},{gao},{keitel}}. The interference in spontaneous
emission occurs when a pair of excited levels of an atom are coupled by the same vacuum modes
to other levels. Many remarkable features have been predicted employing the mechanism of 
interferences in the spontaneous emission of atoms \cite{{gsa},{zhu1},{zhu2},{pas1},{pas2}}. 
The early work of Agarwal on this subject demonstrated trapping of populations in the degenerate 
excited levels of a V-type atom \cite{gsa}. For a non-degenerate V system in free space, 
Zhu {\it et al.} predicted the existence of a dark line in the spontaneous emission 
spectrum \cite{zhu1}. By considering an open V system where the excited atomic levels 
are coupled by a coherent field to another auxiliary level, Scully, Zhu, and coworkers 
showed the possibility of spectral line elimination and spontaneous emission cancellation 
\cite{zhu2} via quantum interference. Phase dependent spectral narrowing \cite{pas1} and  
pulse propagation dynamics \cite{pas2} have also been investigated using the four-level atomic 
model of Ref. \cite{zhu2}.

Since the fluorescence properties of a driven atomic system results from its spontaneous 
emission, studying the influence of interference in such processes has become an important 
topic of research \cite{{cardi},{zhou1},{zhou2},{gao},{keitel}}. The driven V system has been 
shown to exhibit many interference effects such as fluorescence quenching \cite{cardi}, 
ultranarrow spectral lines \cite{zhou1}, anticorrelated photon emissions \cite{zhou2}, 
enhanced squeezing in the fluorescence field \cite{gao}, and collective population trapping 
\cite{keitel}. All these effects assume an existence of non-orthogonal dipole moments of the 
atomic transitions for the interference to occur \cite{gsa}. However, in real atomic systems, 
it is difficult to meet this condition. Different schemes involving cavities with preselected 
polarization \cite{akp1}, coherent- and dc- field induced splitting of atomic levels 
\cite{{akp2},{ficek}} have been proposed later to bypass the condition of non-orthogonal 
dipole moments. Further, the work on spontaneously generated interferences has been extended 
to four level atoms in different configurations. The resonance fluorescence spectrum of driven 
four level atoms in the $\Lambda$-  and V- type configurations has been extensively studied by 
Li {\it et al.} \cite{{fuli1},{fuli2}}. Recently, Ant\'{o}n {\it et al.} \cite{anton} have 
examined a driven four level atom with three excited states and showed that a high population 
inversion may be achieved in the system due to the interference in spontaneous decay channels. 
   
In this paper, we consider a four-level atom in the Y-type scheme (as shown in Fig. 1)
which was proposed earlier for studies on two photon absorption \cite{{bphou},{expt}}. It  
is assumed that the excited atomic states are near-degenerate and decay spontaneously via 
the same vacuum modes to the intermediate state. The atom in the intermediate state can 
further decay to the ground state. Since the cascade decays $(|1\rangle \rightarrow |3\rangle
\rightarrow |4\rangle ~\hbox{and}~ |2\rangle \rightarrow |3\rangle \rightarrow |4\rangle)$ of 
the atom to its ground state from the two initially populated excited states lead to an emission 
of the same pair of photons, the quantum interference exists in decay processes. We investigate 
the role of the interference in the resonance fluorescence from the atom when driven by two 
coherent fields.    

The paper is arranged as follows. In Sec. II, we present the atomic density matrix equations,
describing the interaction of a Y-type atom with two coherent fields, when the presence of 
quantum interference in decay channels is included. The population dynamics of the driven atom in 
the steady state is then studied in Sec. III. In Sec. IV, we analyze the fluorescence spectrum of 
the atom and identify the origin of interference effects using the dressed states of the atomic 
system. Finally, the main results are summarized in Sec. V.

\section{driven Y-type atomic system and its density matrix equations}
We consider (Fig.1) the four-level Y-type atom having two closely lying excited states 
with the energy separation $\hbar W_{12}$. In this scheme, the excited atomic states 
$|1\rangle$ and $|2\rangle$ decay spontaneously to the intermediate state $|3\rangle$ with  
rates  $2 \gamma_1$ and $2 \gamma_2$, respectively. In addition, the atom in the intermediate 
state $|3\rangle$ can undergo spontaneous emissions to the ground state $|4\rangle$ with decay 
rate $2\gamma_3$. We assume that direct transitions between the excited states 
$|1\rangle \rightarrow |2\rangle$ and that between the excited and ground states 
$|1\rangle, |2\rangle \rightarrow |4\rangle$ of the atom are forbidden in the dipole 
approximation. 
\begin{figure}[t]
   \centering
    \includegraphics[width=3.3cm]{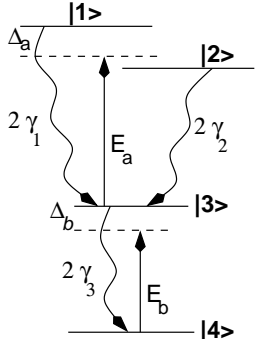}
\caption{The level scheme of the Y-type atom driven by coherent fields.}
\end{figure}
\noindent
A coherent field of frequency $\omega_a$ (amplitude $E_a$)is set to couple the 
upper transitions $|1\rangle, |2\rangle \leftrightarrow |3\rangle$ and another field of
frequency $\omega_b$ (amplitude $E_b$) drives the lower transition 
$|3\rangle \leftrightarrow |4\rangle$. It is further assumed that the transition frequencies
$(\omega_{13},\omega_{23})$ of the upper transitions are widely different from that of the
lower transition $(\omega_{34})$.  The Rabi frequencies of the atom-field interaction
are represented as $\Omega_1 = \vec{\boldsymbol{\mu}}_{13}.\vec{E}_a/\hbar$, 
$\Omega_2 = \vec{\boldsymbol{\mu}}_{23}.\vec{E}_a/\hbar$, and 
$\Omega_3 = \vec{\boldsymbol{\mu}}_{34}.\vec{E}_b/\hbar$ with $\vec{\boldsymbol{\mu}}_{mn}$ being 
the dipole moment of the atomic transition from $|m\rangle$ to $|n\rangle$.
The Hamiltonian of the atom-field interaction is given in the dipole and rotating wave
approximations to be
\begin{eqnarray}
H =&& \hbar \omega_{14} A_{11} + \hbar \omega_{24} A_{22} + \hbar \omega_{34} A_{33}
\nonumber \\
&& - \hbar (\Omega_1 A_{13} e^{-i \omega_a t} + \Omega_2 A_{23} e^{-i \omega_a t} + \hbox{H.c.}) 
\nonumber \\
&&- \hbar(\Omega_3 A_{34} e^{-i \omega_b t} + \hbox{H.c.}).
\end{eqnarray}
Here, the zero of energy is defined at the ground state $|4\rangle$, and $\hbar \omega_{mn}$
is the energy difference between the states $|m\rangle$ and $|n\rangle$. The 
operators $A_{mn} = |m\rangle\langle n|$ represent the atomic population operators for
$m=n$ and transition operators for $m \neq n$. The state $|\psi(t)\rangle$ of the atomic 
system obeys the Schr\"{o}dinger equation 
\begin{equation}
i \hbar \frac{\partial |\psi(t)\rangle}{\partial t} = H |\psi(t)\rangle.
\end{equation}
It is helpful to use the interaction picture by making an unitary transformation
$|\phi\rangle = \exp(i H_0 t/\hbar) |\psi\rangle$ with
\begin{equation}
H_0 = \hbar (\omega_a + \omega_b) A_{11} + \hbar (\omega_a + \omega_b) A_{22} + \hbar \omega_b
A_{33}.
\end{equation}
In the interaction picture, the Schr\"{o}dinger equation for the state $|\phi\rangle$ will
have the effective Hamiltonian given by
\begin{eqnarray}
H_I &=& \hbar (\Delta_a + \Delta_b) A_{11} + \hbar (\Delta_a + \Delta_b - W_{12}) A_{22} 
\nonumber \\
&& + \hbar \Delta_b A_{33} -\hbar (\Omega_1 A_{13} + \Omega_2 A_{23} + \hbox{H.c.}) \nonumber \\ 
&& - \hbar (\Omega_3 A_{34} + \hbox{H.c.}),  \label{ham}
\end{eqnarray}
where $\Delta_a = \omega_{13} - \omega_a$ denotes the detuning between the atomic
frequency $(\omega_{13})$ of the $|1\rangle \rightarrow |3\rangle$ transition 
and the frequency of the applied field $E_a$. Similarly, $\Delta_b = \omega_{34} 
- \omega_b$ corresponds to the detuning of the field applied on the lower transition.  
 
We use the master equation framework to include relaxation processes. With the inclusion
of decay terms, the time evolution of the atomic density matrix describing the atom-field 
interaction obeys
\begin{equation}
\dot{\rho}_{11} = -2 \gamma_1 \rho_{11} + i \Omega_1 (\rho_{31} - \rho_{13})
- p \sqrt{\gamma_1 \gamma_2} (\rho_{12} + \rho_{21}), \label{rho1}
\end{equation}
\begin{equation}  
\dot{\rho}_{22} = -2 \gamma_2 \rho_{22} + i \Omega_2 (\rho_{32} - \rho_{23})
- p \sqrt{\gamma_1 \gamma_2} (\rho_{12} + \rho_{21}), 
\end{equation}
\begin{eqnarray}
\dot{\rho}_{33} =&& 2 \gamma_1 \rho_{11} + 2 \gamma_2 \rho_{22} - 2 \gamma_3 \rho_{33}
+ i \Omega_1 (\rho_{13} - \rho_{31})  \nonumber  \\
&& + i \Omega_2 (\rho_{23} - \rho_{32}) + i \Omega_3 (\rho_{43} - \rho_{34}) \nonumber  \\
&& + 2 p \sqrt{\gamma_1 \gamma_2} (\rho_{12} + \rho_{21}), 
\end{eqnarray}
\begin{eqnarray}
\dot{\rho}_{12} = && - (\gamma_1 + \gamma_2 + i W_{12}) \rho_{12} + i \Omega_1 \rho_{32}
- i \Omega_2 \rho_{13}  \nonumber \\ 
&& - p \sqrt{\gamma_1 \gamma_2} (\rho_{11} + \rho_{22}),
\end{eqnarray}
\begin{eqnarray}
\dot{\rho}_{13} = && - (\gamma_1 + \gamma_3 + i \Delta_a) \rho_{13} + i \Omega_1 
(\rho_{33} - \rho_{11}) - i \Omega_2 \rho_{12} \nonumber \\  
&& - i \Omega_3 \rho_{14} - p \sqrt{\gamma_1 \gamma_2}~ \rho_{23}, 
\end{eqnarray}
\begin{eqnarray}
\dot{\rho}_{23} =&& - [\gamma_2 + \gamma_3 + i (\Delta_a - W_{12})] \rho_{23} +
i \Omega_2 (\rho_{33} - \rho_{22})  \nonumber \\ 
&& - i \Omega_1 \rho_{21} - i \Omega_3 \rho_{24} - p \sqrt{\gamma_1 \gamma_2}~ \rho_{13}, 
\end{eqnarray}
\begin{eqnarray}
\dot{\rho}_{34} = && - (\gamma_3 + i \Delta_b) \rho_{34} + i \Omega_3 (\rho_{44} - \rho_{33})
 + i \Omega_1 \rho_{14} \nonumber \\
&& + i \Omega_2 \rho_{24}, 
\end{eqnarray}
\begin{eqnarray}
\dot{\rho}_{14} = && - [\gamma_1 + i (\Delta_a + \Delta_b)] \rho_{14} + i \Omega_1 \rho_{34}
- i \Omega_3 \rho_{13}  \nonumber \\
&& - p \sqrt{\gamma_1 \gamma_2}~ \rho_{24}, 
\end{eqnarray}
\begin{eqnarray}
\dot{\rho}_{24} = && - [\gamma_2 + i (\Delta_a + \Delta_b - W_{12})] \rho_{24} + i \Omega_2
\rho_{34} - i \Omega_3 \rho_{23}  \nonumber \\
&& - p \sqrt{\gamma_1 \gamma_2}~ \rho_{14},
\label{rho9} 
\end{eqnarray} 
In writing Eqs. (\ref{rho1})-(\ref{rho9}), we have assumed that the trace condition 
$\rho_{11}+\rho_{22} +\rho_{33}+\rho_{44}=1$ is obeyed. The cross-coupling term 
$p \equiv \vec{\boldsymbol{\mu}}_{13}.\vec{\boldsymbol{\mu}}_{23}/|\vec{\boldsymbol{\mu}}_{13}|
|\vec{\boldsymbol{\mu}}_{23}|$ arises due to the quantum interference in spontaneous decay 
transitions. This comes because the decays from the excited states $|1\rangle$ and 
$|2\rangle$ are coupled by the vacuum field. When $p = \pm 1$, the interference effects are 
maximal, whereas if the dipoles are orthogonal $(p=0)$ there is no interference effect in 
spontaneous emission.  

The density matrix equations (\ref{rho1})-(\ref{rho9}) can be rewritten in a more compact
matrix-form by the definition 
\begin{widetext}
\begin{equation}
\hat{\Psi} = {\left(\rho_{11},\rho_{22},\rho_{33},\rho_{12},\rho_{13},\rho_{23},\rho_{14},
\rho_{24},\rho_{34},\rho_{21},\rho_{31},\rho_{32},\rho_{41},\rho_{42},\rho_{43}\right)}^{T}.
\label{psidef}
\end{equation}
\end{widetext}
Substituting Eq. (\ref{psidef}) into Eqs. (\ref{rho1})-(\ref{rho9}), we get the
matrix equation for the variables $\hat{\Psi}_j(t)$
\begin{equation}
\frac{d}{dt} \hat{\Psi} = \hat{L} \hat{\Psi} + \hat{I}, \label{matrix}
\end{equation}
where $\hat{\Psi}_j$ is the $j$-th component of the column vector $\hat{\Psi}$ and the 
inhomogeneous term $\hat{I}$ is also a column vector with non-zero components
\begin{equation}
\hat{I}_9 = i \Omega_3,~~~~~\hat{I}_{15} = -i \Omega_3.
\end{equation} 
In Eq. (\ref{matrix}), $\hat{L}$ is a 15$\times$15 matrix whose elements are
time independent and can be found explicitly from Eqs. (\ref{rho1})-(\ref{rho9}).
The steady state solutions of the density matrix elements can be found by setting the
time derivative equal to zero in Eq. (\ref{matrix}):
\begin{equation}
\hat{\Psi}(\infty) = - \hat{L}^{-1} \hat{I}. \label{steady}
\end{equation}

\section{steady state populations}    
We first study the population dynamics of the driven atom in steady state using
Eq. (\ref{steady}). In Fig. 2, we show the excited and intermediate level
populations $[\rho_{11}(\infty)\equiv\overline{\rho}_{11},\rho_{22}(\infty)\equiv
\overline{\rho}_{22},\rho_{33}(\infty)\equiv\overline{\rho}_{33}]$ versus the detuning 
$\Delta_a$ for different decay rates. All the frequency parameters such as decay rates, 
detuning, and Rabi frequencies are scaled in units of $\gamma_3$. It can be seen in Fig. 2 
that interference effects $(p = 1)$ are less prominent for $\gamma_1, \gamma_2 < \gamma_3$. 
This feature is expected as the interference terms scale as $p\sqrt{\gamma_1 \gamma_2}$ in Eqs. 
(\ref{rho1})-(\ref{rho9}). Further, the graphs show that the excited level populations 
exhibit a resonance at the value of detuning close to $\Delta_a \approx 0$ in the absence of 
interference $(p = 0)$. More generally, the resonances in excited level populations 
$\overline{\rho}_{11}$ and $\overline{\rho}_{22}$ occur when the two photon resonance 
conditions $\Delta_a + \Delta_b = 0$ and $\Delta_a + \Delta_b = W_{12}$ for the $|1\rangle 
\leftrightarrow |4\rangle$ and $|2\rangle \leftrightarrow |4\rangle$ transitions are
respectively satisfied \cite{dalton}. The effect of interference is seen to enhance little 
the population in the excited atomic state when the one photon transitions are resonant, 
$\Delta_a =0,~ \Delta_b = 0$ [see Fig. 2(a)]. Interestingly, for the case of $\gamma_1, 
\gamma_2 \gtrsim \gamma_3$, the interference leads to a splitting of resonances in the excited 
level populations as shown in Fig. 2(b). This result is purely the effect of couplings 
between the different decay pathways that the excited atom can take. It should be borne 
in mind that both one- $(\rho_{13},\rho_{23})$ and two-photon $(\rho_{14},\rho_{24})$ 
coherences contribute in the interference among the decay pathways.
\begin{figure}[ht]
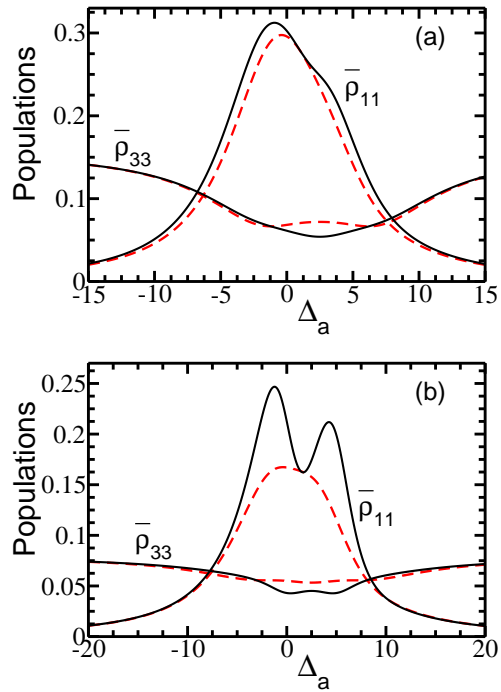

   \vskip 0.4in	
   \includegraphics[width=6.5cm]{fig2a.eps}
   \vskip 0.1in
   \includegraphics[width=6.5cm]{fig2b.eps}	
   \caption{Steady state population of atomic levels as a function of the detuning
   $\Delta_a$ for the parameters $\gamma_3 = 1$, $W_{12} = 5$, $\Delta_b = 0$, 
   $\Omega_1 = \Omega_2 = \Omega_3 = 3$, $\gamma_1 = \gamma_2 = 0.5$ (a), 
   $\gamma_1 = \gamma_2 = 2$ (b). Actual values of $\overline{\rho}_{33}$ are three 
   [six] times than shown in (a) [(b)]. The solid (dashed) curves are for $p = 1$ 
   $(p = 0)$. The curves for $\overline{\rho}_{22}$ (not shown) have a similar behavior 
   as that of $\overline{\rho}_{11}$.}
\end{figure}

To explore further the interference induced splittings of population resonances, we consider
the case of near-degenerate excited levels $(W_{12} \ll \gamma_1,\gamma_2)$ and take the high 
intensity limit $(\Omega_1,\Omega_2,\Omega_3 \gg W_{12},\gamma_1,\gamma_2,\gamma_3)$ of applied 
lasers. For simplicity, we assume equal decay rates $\gamma_1 = \gamma_2 = \gamma$ for the 
upper transitions and examine two different cases, (a) $\gamma \gg \gamma_3$, (b) $\gamma = \gamma_3$,
with respect to the decay rate of the lower transition. The numerical results are shown in Fig. 3
which are to be compared with Fig. 2. It is found that resonances in excited level populations
occur at $\Delta_a = \pm \Omega$ [see Fig. 3(a)] in the limit $\Omega \gg \gamma \gg \gamma_3$, 
where $\Omega_1 = \Omega_2 = \Omega_3 = \Omega$ is considered. In the case of equal decay rates 
$\gamma = \gamma_3$, analytical expressions for the population $(\rho_{11})$ can be obtained
compactly in the presence $(p = 1)$ and absence $(p = 0)$ of interference as
\begin{widetext}
\begin{eqnarray}
\overline{\rho}_{11}(p = 1) &=& \frac{\Omega^4 \Delta_{a}^6 + 12 \Omega^6 \Delta_{a}^4 + 14 \Omega^8 
\Delta_a^2 + 21 \Omega^{10}}{2 \Omega^2 \Delta_a^8 + 16 \Omega^4 \Delta_a^6 + 52 \Omega^6 \Delta_a^4    
 	       + 2 \Omega^8 \Delta_a^2 + 84 \Omega^{10}}, \nonumber \\
\overline{\rho}_{11}(p = 0) &=&	 \frac{4 \Omega^4 \Delta_a^6 + 4 \Omega^8 \Delta_a^4 + 40 \Omega^{10}    
		 \Delta_a^2 + 160 \Omega^{10}}{8 \Omega^2 \Delta_a^8 + 8 \Omega^6 \Delta_a^6 + 64
		 \Omega^8 \Delta_a^4 + 240 \Omega^{10} \Delta_a^2 + 960 \Omega^{10}}, 
		 \label{analy}
\end{eqnarray}		        
\end{widetext}
where all the parameters have been scaled in units of $\gamma$. These analytical formulas account well
for the numerical results in Fig. 3(b). In order to understand physically the effect of interference,
the atomic dynamics is further studied in the bases of symmetric and anti-symmetric states \cite{zhou2}
defined by 
\begin{eqnarray}
|s\rangle &=& \frac{1}{\sqrt{\gamma_1 + \gamma_2}} (\sqrt{\gamma_1} |1\rangle + \sqrt{\gamma_2} |2\rangle),
\nonumber \\
|a\rangle &=& \frac{1}{\sqrt{\gamma_1 + \gamma_2}} (\sqrt{\gamma_2} |1\rangle - \sqrt{\gamma_1} |2\rangle).
\label{couple}
\end{eqnarray}
\begin{figure}[b]
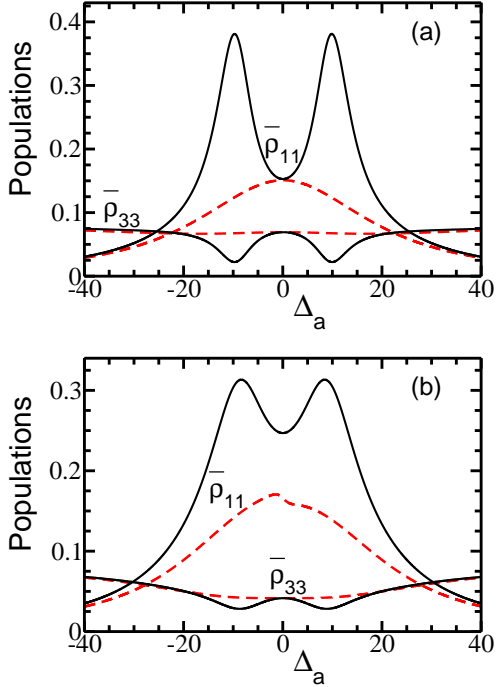

   \includegraphics[width=6.5cm]{fig3a.eps}
   \vskip 0.1in
   \includegraphics[width=6.5cm]{fig3b.eps}	
    \caption{Steady state population of atomic levels as a function of the detuning
   $\Delta_a$ for the parameters $\gamma_3 = 1$, $W_{12} = 0.2$, $\Delta_b = 0$,
   $\Omega_1 = \Omega_2 = \Omega_3 = 10$, $\gamma_1 = \gamma_2 = 5$ (a),
   $\gamma_1 = \gamma_2 = 1$ (b). Actual values of $\overline{\rho}_{33}$ are six
   times than shown. The solid (dashed) curves are for $p = 1$ $(p = 0)$.}
\end{figure}
With $\gamma_1 = \gamma_2 = \gamma$ and using Eq. (\ref{couple}), the Hamiltonian Eq. (\ref{ham}) can be 
rewritten as
\begin{eqnarray}
H_I = && \hbar (\Delta_a - \frac{W_{12}}{2}) (|s\rangle \langle s| + |a\rangle \langle a|) \nonumber \\
&& + \frac{\hbar W_{12}}{2} (|s\rangle \langle a| + |a\rangle \langle s|) 
 - \hbar \sqrt{2} \Omega (|s\rangle \langle 3| + |3\rangle \langle s|)  \nonumber \\
&& -\hbar \Omega (|3\rangle \langle 4| + |4\rangle \langle 3|). 
\end{eqnarray}
From the above Hamiltonian, it is seen that only the symmetric state $|s\rangle$ is interacting with the 
light field. However, the antisymmetric state $|a\rangle$ may be populated by its coupling with the symmetric 
state because of the separation energy $(\hbar W_{12})$ between the excited atomic levels. This can become 
clear by analyzing the density matrix equations in the bases (\ref{couple}) :
\begin{eqnarray}
\dot{\rho}_{aa} &=& \frac{i W_{12}}{2} \rho_{as} - \frac{i W_{12}}{2} \rho_{sa}, \nonumber \\
\dot{\rho}_{ss} &=& - 4 \gamma \rho_{ss} - \frac{i W_{12}}{2} \rho_{as} + \frac{i W_{12}}{2} \rho_{sa}
\nonumber \\
&&~~+ i \sqrt{2} \Omega (\rho_{3s} - \rho_{s3}).
\end{eqnarray}
Here, the case of maximal quantum interference $(p = 1)$ has been assumed. It is evident that the 
antisymmetric state is a non-decaying state for $p = 1$ and it is coupled to the symmetric state for 
$W_{12} \neq 0$ (though small as in Fig. 3). In Fig. 4, the steady state populations of the symmetric 
and antisymmetric states are plotted for the same parameters of Fig. 3(a). The graphs show that the 
splitting of resonances occurs due to a high population of the antisymmetric state. We have so far 
assumed a fixed value for the lower transition detuning $(\Delta_b = 0)$ and studied the dependence of 
populations on the upper transition detuning $(\Delta_a)$. However, the results (not shown) will be 
qualitatively similar even in the general case of varying both $\Delta_a$ and $\Delta_b$.        
\begin{figure}[htb]	
   \centering
   \includegraphics[width=6.5cm]{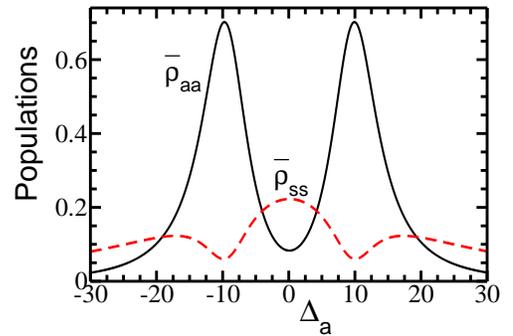}	
    \caption{Steady state populations $\overline{\rho}_{aa}$ and $\overline{\rho}_{ss}$
    as a function of the detuning $\Delta_a$ for the same parameters of Fig. 3(a) with
    $p = 1$.}
\end{figure}
\section{Resonance Fluorescence Spectrum}
We now proceed to the study of the resonance fluorescence spectra of the driven atom. Since 
the atom is driven by two coherent fields, each field induces its own atomic dipole moment which then
generates a scattered field. However, the fields scattered by the upper- and lower- transitions in the 
atom will have no correlations because the applied fields ($E_a$ ,$E_b$) are of quite different 
carrier frequencies ($\omega_a$, $\omega_b$). In the interaction picture, the negative- and 
positive-frequency parts of the polarization operators are written as   
\begin{figure}[t]
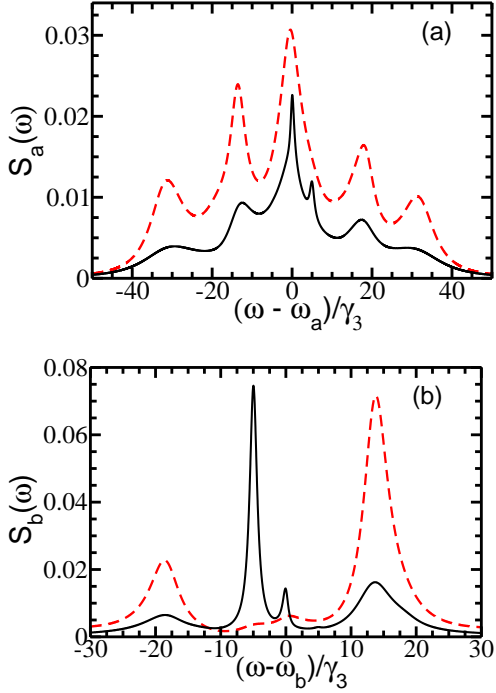

   \includegraphics[width=6.5cm]{fig5a.eps}
   \vskip 0.1in
   \includegraphics[width=6.5cm]{fig5b.eps}	
    \caption{The incoherent spectrum of the fluorescent field generated by the dipoles 
    (a) $\hat{\mathbf P}^{(+)}_{\omega_a}$ and (b) $\hat{\mathbf P}^{(+)}_{\omega_b}$ for the
    parameters $\gamma_3 = 1$, $W_{12} = 10$, $\Delta_a = \Delta_b = 0$, $\Omega_1 = \Omega_2
    = 10$, $\Omega_3 = 5$, $\gamma_1 = \gamma_2 = 3$. The solid (dashed) curves are for $p = 1$ 
    $(p = 0)$.} 
\end{figure}
\begin{eqnarray}
\hat{\mathbf P}^{(-)}_{\omega_a}(t) &=& \vec{\boldsymbol{\mu}}_{13} e^{i\omega_a t} |1\rangle\langle 3|
+ \vec{\boldsymbol{\mu}}_{23} e^{i\omega_a t} |2\rangle\langle 3|, \nonumber \\
\hat{\mathbf P}^{(+)}_{\omega_a}(t) &=&  [\hat{\mathbf P}^{(-)}_{\omega_a}(t)]^{\dagger},
\end{eqnarray} 

\begin{equation}
\hat{\mathbf P}^{(-)}_{\omega_b}(t) =  \vec{\boldsymbol{\mu}}_{34} e^{i \omega_b t} |3\rangle\langle 4|,
~~~~~~
\hat{\mathbf P}^{(+)}_{\omega_b}(t) = [\hat{\mathbf P}^{(-)}_{\omega_b}(t)]^{\dagger}.
\end{equation} 
To calculate the fluorescence spectra, we need the two-time expectation values of the polarization
operator. The spectrum of resonance fluorescence is defined by the Fourier transformation of the
two-time correlation or equivalently the real part of its Laplace transform:
\begin{eqnarray}
S_{a}(\omega) = \hbox{Re} \int_{0}^{\infty} \lim_{t \rightarrow \infty}
\langle \hat{\mathbf P}^{(-)}_{\omega_{a}}(t+\tau).\hat{\mathbf P}^{(+)}_{\omega_{a}}(t)
\rangle  e^{-i \omega \tau} d\tau, &&\nonumber  \\
S_{b}(\omega)  = \hbox{Re} \int_{0}^{\infty} \lim_{t \rightarrow \infty}
\langle \hat{\mathbf P}^{(-)}_{\omega_{b}}(t+\tau).\hat{\mathbf P}^{(+)}_{\omega_{b}}(t)
\rangle  e^{-i \omega \tau} d\tau.&&  \label{spect}
\end{eqnarray}
Here, the index $a$ $(b)$ refers to the spectrum of the fluorescence light emitted by
the atom with a central frequency $\omega_a$ $(\omega_b)$. The Laplace transformation
with variable $Z = i \omega$ of the correlation function, defined in the spectrum above,
has a pole at $Z = i \omega_a$ $(Z = i \omega_b)$ which attributes to the coherent 
Rayleigh scattering of the spectrum. The incoherent part is obtained by removing the
contributions of the poles. 

With the application of the quantum regression theorem \cite{{fuli1},{fuli2},{narducci}}
and using the steady state solutions Eq. (\ref{steady}) of the density matrix elements,
the incoherent fluorescence spectra can be obtained as
\begin{eqnarray}
S_{a}(\omega) &=& \hbox{Re}\Big\{ {|\vec{\boldsymbol{\mu}}_{13}|}^2 \big[\hat{M}_{11,9} 
\overline{\rho}_{14} + \hat{M}_{11,3} \overline{\rho}_{13} + \hat{M}_{11,12}
\overline{\rho}_{12} \nonumber \\
&& + \hat{M}_{11,11} \overline{\rho}_{11} + \sum_{j=1}^{15} \hat{N}_{11,j} \hat{I}_j 
\overline{\rho}_{13} \big] +\vec{\boldsymbol{\mu}}_{23}.\vec{\boldsymbol{\mu}}_{13}^{*} 
\nonumber \\
&& \times  \big[\hat{M}_{12,9} \overline{\rho}_{14} + \hat{M}_{12,3} \overline{\rho}_{13} 
+ \hat{M}_{12,12} \overline{\rho}_{12} + \hat{M}_{12,11} \nonumber \\
&&  \times \overline{\rho}_{11} + \sum_{j=1}^{15} \hat{N}_{12,j}\hat{I}_j \overline{\rho}_{13} \big]  
+ \vec{\boldsymbol{\mu}}_{13}.\vec{\boldsymbol{\mu}}_{23}^{*} 
\big[\hat{M}_{11,11} \overline{\rho}_{21} \nonumber \\
&& + \hat{M}_{11,9} \overline{\rho}_{24} + \hat{M}_{11,3} \overline{\rho}_{23} + \hat{M}_{11,12} 
\overline{\rho}_{22} \label{specta} \\
&& + \sum_{j=1}^{15} \hat{N}_{11,j} \hat{I}_j \overline{\rho}_{23} \big] + 
{|\vec{\boldsymbol{\mu}}_{23}|}^2 \big[ \hat{M}_{12,11} \overline{\rho}_{21} 
+ \hat{M}_{12,9} \nonumber \\
&& \times \overline{\rho}_{24} + \hat{M}_{12,3} \overline{\rho}_{23} + \hat{M}_{12,12} 
\overline{\rho}_{22} + \sum_{j=1}^{15} \hat{N}_{12,j} \hat{I}_j \nonumber \\ 
&& \times \overline{\rho}_{23} \big] \Big\}, \nonumber                                                                                         
\end{eqnarray}
where the matrices $\hat{M} = {(Z - \hat{L})}^{-1}|_{Z = i (\omega - \omega_a)}$ and
$\hat{N} = \hat{L}^{-1}\hat{M}$. Similarly, 
\begin{eqnarray}
S_{b}(\omega) = && \hbox{Re}\Big\{ {|\vec{\boldsymbol{\mu}}_{34}|}^2 \big[ \hat{M}_{15,13}
\overline{\rho}_{31} + \hat{M}_{15,14} \overline{\rho}_{32} + \hat{M}_{15,15} 
\overline{\rho}_{33} \nonumber \\
&& + \sum_{j=1}^{15} \hat{N}_{15,j} \hat{I}_j \overline{\rho}_{34} \big]\Big \}, 
\label{spectb}
\end{eqnarray}
with the matrices $\hat{M} = {(Z - \hat{L})}^{-1}|_{Z = i (\omega - \omega_b)}$ and 
$\hat{N} = \hat{L}^{-1}\hat{M}$. 

The set of equations (\ref{specta}) and (\ref{spectb}) can be used to obtain numerically
the spectral characteristics of the driven atom. Figure 5 displays the numerical results by 
assuming equal decay rates $\gamma_1 = \gamma_2$ of the upper atomic transitions. The spectra 
$S_a(\omega)$ and $S_b(\omega)$ are scaled in units of $|\vec{\boldsymbol{\mu}}_{13}|^2 \gamma_3^{-1}$ 
and $|\vec{\boldsymbol{\mu}}_{34}|^2 \gamma_3^{-1}$ respectively. In the presence of quantum
interference $(p = 1)$, the spectrum shows the typical line narrowing effect, as discussed
in earlier publications \cite{{zhou1},{fuli1},{fuli2}}, in the fluorescent field with the central 
frequency $\omega_a$ [see Fig. 5(a)]. However, the spectral features get remarkably modified in the 
fluorescent field emitted by the lower atomic transitions. It is seen that the inner sideband in the 
fluorescence spectrum gets enhanced due to interference with a corresponding reduction in the intensity 
of the outer sidebands [compare solid and dashed curves in Fig. 5(b)]. A physical understanding of 
this interesting result can be obtained in the dressed state description of the atom-field interaction. 
The dressed states are defined as eigenstates of the time independent Hamiltonian (\ref{ham}) :
\begin{equation}
H_I |\Phi\rangle = \hbar \lambda|\Phi\rangle. \label{ham2}
\end{equation} 
\begin{figure}[t]
   \centering
   \includegraphics[width=6.5cm]{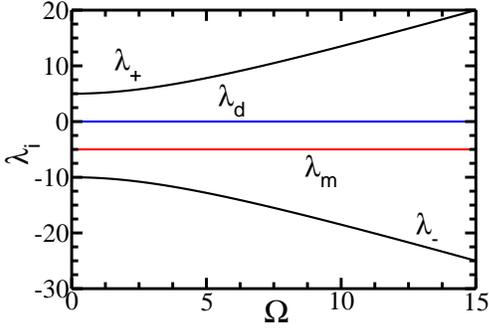}	
    \caption{The dressed state eigenvalues versus the Rabi frequency $\Omega$ for
    the parameters $\gamma_3 = 1$, $W_{12} = 10$, $\Delta_a = \Delta_b = 0$, $\Omega_3 = 5$.}   
\end{figure}
\noindent    
In the general parametric conditions, it is difficult to find analytical solutions to the 
eigenvalue equation (\ref{ham2}). For simplicity, the case of two photon resonance 
$\Delta_a = \Delta_b = 0$ is assumed in the following. In this case, there exists an eigenstate 
$|d\rangle$ with the eigenvalue $\lambda_d = 0$ as
\begin{equation}
|d\rangle = \frac{1}{\sqrt{\Omega_1^2 + \Omega_3^2}} [\Omega_3 |1\rangle - \Omega_1 |4\rangle].
\label{dark}
\end{equation} 
The non-zero eigenvalues and the corresponding eigenstates can be obtained by diagonalizing the 
Hamiltonian $H_I$ in the basis of bare atomic states. We consider a special choice of 
parameters $\Omega_1 = \Omega_2 = \Omega$ and $\Omega_3 = W_{12}/2$ [as in Fig. 5] which
allows for simple analytical solutions. The operator $H_I$ has eigenstates $|m\rangle$,
$|\pm\rangle$ with eigenvalues (in units of $\hbar$) $\lambda_m = - W_{12}/2$, 
$\lambda_{\pm} = (- W_{12} \pm \sqrt{W_{12}^2 + 32 (\Omega^2 + \frac{W_{12}^2}{4})})/4$, respectively, 
where  
\begin{eqnarray}
|m\rangle &=& \frac{1}{\sqrt{2(\Omega^2 + (W_{12}^2/4))}} \nonumber  \\
&&\times \left[\Omega |1\rangle - \Omega |2\rangle + \frac{W_{12}}{2} |3\rangle + \frac{W_{12}}{2} 
|4\rangle \right], \\
|\pm\rangle &=& N_{\pm} \left[\Omega |1\rangle + \frac{\lambda_{\pm} \Omega}{(W_{12} + \lambda_{\pm})} 
|2\rangle - \lambda_{\pm} |3\rangle + \frac{W_{12}}{2} |4\rangle \right], \nonumber \label{dress} 
\end{eqnarray}
with $N_{\pm} = 1\Big/\sqrt{\lambda_{\pm}^2 [1 + \Omega^2/{(W_{12} + \lambda_{\pm})}^2] + \Omega^2 +
W_{12}^2/4}$. 

In order to interpret the numerical results in Fig. 5, we study the behavior of the 
dressed states in steady state with the inclusion of decay processes using Eq. (\ref{steady}).
In Figs. 6 and 7, the dressed state eigenvalues $(\lambda \hbox{values})$ and its populations are shown 
as a function of the Rabi frequency $\Omega$ for the fixed value of $\Omega_3 = W_{12}/2$. Note that the 
eigenvalues $\lambda_d$ and $\lambda_m$ are independent of the parameter $\Omega$. The peaks in the 
fluorescence spectrum can be attributed to transitions between the dressed states 
$|\Phi'\rangle \leftrightarrow |\Phi \rangle$ $(\Phi,\Phi' = d,m,+,-)$. For $p = 0$ and 
$\Omega \gtrsim W_{12}$, the dressed states $|d\rangle$, $|+\rangle$, and $|-\rangle$ are well 
populated as shown in Fig. 7. 
\begin{figure}[t]
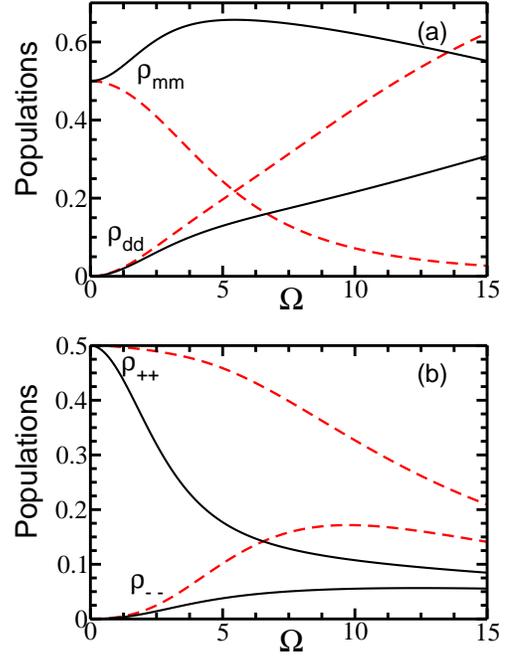

   \includegraphics[width=6.5cm]{fig7a.eps}
   \vskip 0.1in
   \includegraphics[width=6.5cm]{fig7b.eps}	
    \caption{Steady state population of dressed states, (a) $\rho_{mm}$, $\rho_{dd}$
    and (b) $\rho_{++}$, $\rho_{--}$, as a function of the Rabi frequency $\Omega$ for the 
    parameters $\gamma_3 = 1$, $W_{12} = 10$, $\Delta_a = \Delta_b = 0$, $\Omega_3 = 5$, 
    $\gamma_1 = \gamma_2 = 3$. The solid (dashed) curves are for $p = 1$ $(p = 0)$.}
\end{figure}
\noindent      
The fluorescence peaks in Fig. 5 occur at the energy differences 
between these states. However, in the presence of interference $(p = 1)$, the atomic population 
accumulates mostly in the dressed state $|m\rangle$ [see Fig. 7(a)]. This can be explained as due 
to a destructive quantum interference among the spontaneous decay pathways. The rate of transitions 
$|\Phi\rangle \rightarrow |\Phi'\rangle$ between dressed states $|\Phi\rangle$ and $|\Phi'\rangle$ 
is given by the squared dipole matrix elements which for the emission lines with the central 
frequencies $\omega_a$ and $\omega_b$ becomes
\begin{eqnarray}  
R_{\Phi,\Phi'}^{a} &=& {|\langle \Phi'|\hat{\mathbf P}^{(+)}_{\omega_a}|\Phi\rangle|}^{2} \nonumber \\
&=& |\vec{\boldsymbol{\mu}}_{13}|^2 [C_{1\Phi}^2 + C_{2\Phi}^2 + 2 p C_{1\Phi} C_{2\Phi}] C_{3\Phi'}^2,
\label{rate1} \\
R_{\Phi,\Phi'}^{b} &=& {|\langle \Phi'|\hat{\mathbf P}^{(+)}_{\omega_b}|\Phi\rangle|}^{2} \nonumber \\
&=& |\vec{\boldsymbol{\mu}}_{34}|^2 C_{3\Phi}^2 C_{4\Phi'}^2,  \label{rate2}
\end{eqnarray}
where $|\vec{\boldsymbol{\mu}}_{13}| = |\vec{\boldsymbol{\mu}}_{23}|$ has been assumed and $C_{i\Phi}$'s
$[i = 1,2,3,4]$ denote the coefficients of the bare atomic states $|i\rangle$ in the dressed state 
$|\Phi\rangle$. As seen in the above Eq. (\ref{rate1}), there is an interference term (p term) in the 
transition rate $R_{\Phi,\Phi'}^{a}$ as a result of spontaneous decays along the $|1\rangle 
\rightarrow |3\rangle$ and $|2\rangle \rightarrow |3\rangle$ transitions. For the dressed state 
$|m\rangle$, this term cancels the square factors in the case of maximal interference $(p = 1)$, thus 
suppressing the atomic decay, narrowing the spectral lines [shown in Fig. 5(a)] and enhancing the 
population [shown in Fig. 7(a)] in this state. However, the dressed state $|m\rangle$ can decay because 
of spontaneous emissions along the $|3\rangle \rightarrow |4\rangle$ transitions even when $p = 1$ 
[see Eq. (\ref{rate2})]. This leads to the enhancement of the inner sideband in the spectrum 
[shown in Fig. 5(b)] of the fluorescence light emitted by the lower transitions in the atom. It is 
because only the states $|m\rangle$ and $|d\rangle$ are populated mainly in steady state. Finally, we note that
the existence of atomic steady state and discussions so far assume the non-degenerate $(W_{12} \neq 0)$
case of excited atomic levels. In the degenerate case $(W_{12} = 0)$, there exists no unique solution
to Eq. (\ref{matrix}) in steady state. In fact, the steady state fluorescence properties become dependent 
on the initial conditions due to degeneracy of the dressed states of the Hamiltonian.  

\vspace{2cm}
\section{summary}
We have investigated the resonance fluorescence from a driven Y-type atom when the 
presence of interference in spontaneous decay channels is important. At first, the 
steady state dynamics of the atom was studied using the density matrix approach. 
We have shown that the decay-induced interference can lead to splitting of resonances 
in the excited level populations calculated as a function of light detuning. This has 
been explained as due to high population of a non-decaying anti-symmetric state of the 
atom. Then, the role of interference in the spectral characteristics of the driven
atom was examined. It is found that the interference results in narrowing of central
peaks and enhancement of inner sidebands in the fluorescence spectrum. A physical
understanding of the numerical results has been presented based on the dressed state
theory of atom-field interaction. Clearly, the present work is open ended with the effects
of interference in driven Y systems on two photon correlations and squeezing 
spectra remaining unexplored. Detailed investigations of such studies will be published 
elsewhere.

\vspace{-1cm}
\begin{acknowledgments}
The author thanks Prof. G.S. Agarwal for useful suggestions and encouragements.
\end{acknowledgments}

\end{document}